\input harvmac
\skip0=\baselineskip
\divide\skip0 by 2

\noblackbox
\lref\ckl{hep-th/0105227 }

\lref\pope{Pope et al}

\lref\SenTT{
A.~Sen,
``SO(32) spinors of type I and other solitons on brane-antibrane pair,''
JHEP {\bf 9809}, 023 (1998)
[hep-th/9808141].
}

\lref\SenMG{
A.~Sen,
``Non-BPS states and branes in string theory,''
hep-th/9904207.
}
\lref\taub{C. Taubes ``The Existence of a Nonminimal Solution to the
Yang-Mills Higgs Equations on R3'', Comm. Math. Phys. {\bf 86} (1982) 257;
{\bf 86} (1982) 299.}
\lref\manu{J.~M.~Maldacena and C.~Nunez,
``Towards the large N limit of pure N = 1 super Yang Mills,''
Phys.\ Rev.\ Lett.\  {\bf 86}, 588 (2001), [hep-th/0008001].}
\lref\CohenPV{
A.~G.~Cohen, G.~W.~Moore, P.~Nelson and J.~Polchinski,
``Semi Off-Shell String Amplitudes,''
Nucl.\ Phys.\ B {\bf 281}, 127 (1987).
}

\lref\PolchinskiMT{
J.~Polchinski,
``Dirichlet-Branes and Ramond-Ramond Charges,''
Phys.\ Rev.\ Lett.\  {\bf 75}, 4724 (1995)
[hep-th/9510017].
}

\lref\GreenWR{
M.~B.~Green and J.~A.~Shapiro,
``Off-Shell States In The Dual Model,''
Phys.\ Lett.\ B {\bf 64}, 454 (1976).
}
\lref\Bala{
V.~Balasubramanian, J.~de Boer and D.~Minic,
``Mass, entropy and holography in asymptotically de Sitter spaces,''
hep-th/0110108.}
\lref\GreenZV{
M.~B.~Green,
``Pointlike Structure And Off-Shell Dual Strings,''
Nucl.\ Phys.\ B {\bf 124}, 461 (1977).
}

\lref\joe{J.~Polchinski,
``Dirichlet-Branes and Ramond-Ramond Charges,''
Phys.\ Rev.\ Lett.\  {\bf 75}, 4724 (1995)
[hep-th/9510017].}

\lref\GreenYT{
M.~B.~Green,
``Dynamical Pointlike Structure And Dual Strings,''
Phys.\ Lett.\ B {\bf 69}, 89 (1977).
}

\lref\GreenIW{
M.~B.~Green,
``Pointlike States For Type 2b Superstrings,''
Phys.\ Lett.\ B {\bf 329}, 435 (1994)
[hep-th/9403040].
}

\lref\ckl{
B.~Craps, P.~Kraus and F.~Larsen,
``Loop corrected tachyon condensation,''
JHEP {\bf 0106}, 062 (2001)
[hep-th/0105227].
}

\lref\FarhiEH{
E.~Farhi, J.~Goldstone, S.~Gutmann, K.~Rajagopal and R.~J.~Singleton,
``Fermion production in the background of Minkowski space classical solutions in spontaneously broken gauge theory,''
Phys.\ Rev.\ D {\bf 51}, 4561 (1995)
[hep-ph/9410365].
}

\lref\DiVecchiaPR{
P.~Di Vecchia, M.~Frau, I.~Pesando, S.~Sciuto, A.~Lerda and R.~Russo,
``Classical p-branes from boundary state,''
Nucl.\ Phys.\ B {\bf 507}, 259 (1997)
[hep-th/9707068],~~P.~Di Vecchia and A.~Liccardo,
``D-branes in string theory. II,''
hep-th/9912275.

}

\lref\GreenUM{
M.~B.~Green and M.~Gutperle,
``Light-cone supersymmetry and D-branes,''
Nucl.\ Phys.\ B {\bf 476}, 484 (1996)
[hep-th/9604091].
}

\lref\GubserBC{
S.~S.~Gubser, I.~R.~Klebanov and A.~M.~Polyakov,
``Gauge theory correlators from non-critical string theory,''
Phys.\ Lett.\ B {\bf 428}, 105 (1998)
[hep-th/9802109].
}

\lref\MaldacenaRE{
J.~Maldacena,
``The large $N$ limit of superconformal field theories and supergravity,''
Adv.\ Theor.\ Math.\ Phys.\  {\bf 2}, 231 (1998)
[Int.\ J.\ Theor.\ Phys.\  {\bf 38}, 1113 (1998)]
[hep-th/9711200].
}

\lref\WittenQJ{
E.~Witten,
``Anti-de Sitter space and holography,''
Adv.\ Theor.\ Math.\ Phys.\  {\bf 2}, 253 (1998)
[hep-th/9802150].
}

\lref\StromingerP{
A.~Strominger,
``The dS/CFT correspondence,''
JHEP {\bf 0110}, 034 (2001)
[hep-th/0106113].
}

\lref\StromingerGP{
A.~Strominger,
``Inflation and the dS/CFT correspondence,''
JHEP {\bf 0111}, 049 (2001)
[hep-th/0110087].
}

\lref\DouglasBN{
M.~R.~Douglas,
``Branes within branes,''
hep-th/9512077.
}

\lref\HullVG{
C.~M.~Hull,
``Timelike T-duality, de Sitter space, large N gauge theories and  topological field theory,''
JHEP {\bf 9807}, 021 (1998)
[hep-th/9806146].
}

\lref\HullII{
C.~M.~Hull,
``de Sitter space in supergravity and M theory,''
JHEP {\bf 0111}, 012 (2001)
[hep-th/0109213].
}
\lref\vijay{V. Balasubramanian, S. F. Hassan, E. Keski-Vakkuri,
and A. Naqvi ``A Space-Time Orbifold: A Toy Model for a
Cosmological Singularity'',  hep-th/0202187.}

\lref\lms{H. Liu, G. Moore and N. Seiberg, to appear.}

\lref\HarveyQU{
J.~A.~Harvey, P.~Horava and P.~Kraus,
``D-sphalerons and the topology of string configuration space,''
JHEP {\bf 0003}, 021 (2000)
[hep-th/0001143].
}

\lref\PolchinskiFQ{
J.~Polchinski,
``Combinatorics Of Boundaries In String Theory,''
Phys.\ Rev.\ D {\bf 50}, 6041 (1994)
[hep-th/9407031].
}

\lref\GreenTV{
M.~B.~Green and M.~Gutperle,
``Effects of D-instantons,''
Nucl.\ Phys.\ B {\bf 498}, 195 (1997)
[hep-th/9701093].
}

\lref\GreenMY{
M.~B.~Green,
``A Gas of D instantons,''
Phys.\ Lett.\ B {\bf 354}, 271 (1995)
[hep-th/9504108].
}
\lref\MantonND{
N.~S.~Manton,
``Topology In The Weinberg-Salam Theory,''
Phys.\ Rev.\ D {\bf 28}, 2019 (1983).
}

\lref\KlinkhamerDI{
F.~R.~Klinkhamer and N.~S.~Manton,
``A Saddle Point Solution In The Weinberg-Salam Theory,''
Phys.\ Rev.\ D {\bf 30}, 2212 (1984).
}

\lref\KoganNN{
I.~I.~Kogan and N.~B.~Reis,
``H-branes and chiral strings,''
Int.\ J.\ Mod.\ Phys.\ A {\bf 16}, 4567 (2001)
[arXiv:hep-th/0107163].
}

\lref\LuJK{ H.~Lu, S.~Mukherji, C.~N.~Pope and K.~W.~Xu,
``Cosmological solutions in string theories,'' Phys.\ Rev.\ D {\bf
55}, 7926 (1997) [arXiv:hep-th/9610107].
}

\lref\LuER{ H.~Lu, S.~Mukherji and C.~N.~Pope,
 ``From p-branes to
cosmology,'' Int.\ J.\ Mod.\ Phys.\ A {\bf 14}, 4121 (1999)
[arXiv:hep-th/9612224].
}

\lref\LukasEE{ A.~Lukas, B.~A.~Ovrut and D.~Waldram,
``Cosmological solutions of type II string theory,'' Phys.\ Lett.\
B {\bf 393}, 65 (1997) [arXiv:hep-th/9608195].
}

\lref\LukasIQ{ A.~Lukas, B.~A.~Ovrut and D.~Waldram, ``String and
M-theory cosmological solutions with Ramond forms,'' Nucl.\ Phys.\
B {\bf 495}, 365 (1997) [arXiv:hep-th/9610238].
}

\lref\Quevedo{C.~Grojean, F.~Quevedo, G.~Tasinato and  I. Zavala, ``Branes
 on Charged Dilatonic Backgrounds: Self-Tuning, Lorentz Violations and 
Cosmology,'' JHEP {\bf 0108}, 005 (2001) [arXiv:hep-th/0106120].
}

\def\t{\tau}

\def\l{\lambda}
\def\p{\partial}
\def\msurr{\mathsurround=0pt}
\def\overleftrightarrow#1{\vbox{\msurr\ialign{##\crcr
        $\leftrightarrow$\crcr\noalign{\kern-1pt\nointerlineskip}
        $\hfil\displaystyle{#1}\hfil$\crcr}}}

\Title{\vbox{\baselineskip12pt\hbox{hep-th/0202210}
\hbox{HUTP-02/A001}
\hbox{}}}{Spacelike Branes}

\centerline{Michael Gutperle and Andrew Strominger}
\bigskip\centerline{Jefferson Physical Laboratory}
\centerline{Harvard University} \centerline{Cambridge, MA 02138}
\vskip .3in \centerline{\bf Abstract}
{ Scalar field theories  with appropriate potentials
in Minkowski space can have time-dependent
classical solutions containing topological defects which correspond to
S-branes - $i.e.$ branes all of whose tangential dimensions are spacelike.
It is argued that such S-branes arise in string theory as time-dependent
 solutions of the worldvolume tachyon field of an unstable D-brane or
D-brane-anti-D-brane pair. Using the known coupling of the
spacetime RR fields to the worldvolume tachyon it is shown that
these S-branes carry a charge, defined as the integral of a RR
field strength over a sphere (containing a time as well as spatial
dimensions) surrounding the S-brane. This same charge is carried by
SD-branes, $i.e.$ Dirichlet branes arising from
open string worldsheet conformal
field theories with a Dirichlet boundary condition on the timelike
dimension. The corresponding SD-brane boundary
state is constructed. Supergravity
solutions carrying the same charges are also found for a few cases.   }

\smallskip
\Date{}
\listtoc
\writetoc
\newsec{Introduction}

In this paper it is argued that string theory - as well as many field
theories - contain time dependent
solutions corresponding to spacelike branes or S-branes.
These are topological defects which are localized on a
spacelike hypersurface and hence exist only for a moment  in time.
It is further suggested that some of these defects
are related to spacelike versions of
the familiar D-branes \joe\ (or SD-branes), in which the
time coordinate obeys a
Dirichlet boundary condition. The main piece of evidence for this
proposed relation is that they
carry the same type of RR charge.  However, as discussed herein,
a detailed analysis
indicates possible subtleties in the relationship.
Similar arguments suggest the existence of branes on null hypersurfaces or
N-branes.

Interest in such configurations arises from several related viewpoints.
Over the past several decades an astonishing and rich variety of
phenomena have been discovered in string theory which have altered our
fundamental
notion of space. Virtually all of these phenomena
concern static geometries. In general relativity, space and time are on
the same footing, so we should expect equally surprising phenomena in the
time-dependent context.  In order to find these phenomena we should study
time-dependent string backgrounds.   Flat spacetimes with a Dirichlet boundary
condition in the time direction (an SD-brane)  or null direction
(ND-brane) are some of the
simplest imaginable time dependent conformal field theories.

One of the concepts one hopes to generalize from a spatial to a temporal
context is holography. In the AdS/CFT correspondence
\refs{\MaldacenaRE\GubserBC - \WittenQJ }, the
D-brane field theory
holographically reconstructs a spatial dimension. By analogy, SD-branes
should  holographically reconstruct a time dimension. Such a
temporal reconstruction
was argued to be a key ingredient of a proposed dS/CFT correspondence
\refs{ \StromingerP\StromingerGP-\Bala} (see also \refs{\HullVG,\HullII}).
Indeed the present
 work was in large part motivated in an attempt to understand how
string theory might produce the Euclidean field theories required for
a dS/CFT correspondence.  S-branes might also be relevant for the
resolution of the spacelike singularities inside black holes.

The basic idea - illustrated by a specific example -
is as follows. Type IIA string theory contains
an unstable D3-brane with a tachyon field whose potential
resembles a double well. Sen \SenMG\ has argued the the stable D2-brane
is the tachyonic kink solution of the unstable D3 worldvolume field
theory. Now consider initial data at  time $t=0$ for the D3 with the
tachyon field perched at the unstable minimum and given a small
constant positive velocity. As time evolves into the future, the tachyon
will roll off the top, emit closed string radiation, and eventually settle
in to the positive minimum. Evolving into the past
from $t=0$, one finds the time
reversed picture with the tachyon approaching the negative minimum at
$t=-\infty$. The full picture consists of
finely tuned incoming radiation which conspires to excite the
tachyon field to the top of the potential barrier. The tachyon then rolls
down the other side and dissipates its energy back in to radiation.
The result is a timelike kink in the tachyon field which is an
S2-brane.\foot{We adopt conventions such that
an Sp-brane has p+1 spatial dimensions and no time
dimensions.}
A similar discussion involving initial data on null hypersurfaces leads to
N-branes (see \KoganNN).

Using the known couplings of the open string tachyon to the RR fields,
one concludes that this configuration carries the same kind of
charge\foot{This charge is measured by an integral of the RR 6-form over
the $S^6$ surrounding the brane. This $S^6$ has both timelike and spacelike
dimensions. It is perhaps an abuse of terminology to refer to such an
integral as a charge because it is not time independent, but we
nevertheless find it convenient. } as that carried by a spacelike
D2-brane. In analogy with Sen's identification of the spacelike
kink as an ordinary D2-brane, this motivates the identification of
the timelike kink as an SD2-brane.

This construction has obvious generalizations to other
(co)dimensions, as well as to branes as vortices in
brane-antibrane pairs, etc. Furthermore in analogy to the
description of branes within branes as instantons \DouglasBN,
S-branes within branes can be described as decaying sphaleron type
configurations. In fact in line with the discussion of \HarveyQU\
this might be related to the tachyon soliton description of
branes.

 In analogy with AdS/CFT, one anticipates that the worldvolume
field theory on a stack of $N$ S-branes is holographically dual to
the supergravity solution sourced by the S-branes. As a small step
in investigating such a possible connection, we find two examples
of such spacetime solutions.

This paper is organized as follows. In section 2.1 a real scalar
field in four spacetime dimensions with a double well potential is
considered, and argued to have a timelike kink which corresponds
to an S2-brane.  Section 2.2 considers a complex scalar with a
Mexican hat potential and describes the S1-brane solution. In 2.3
three-form and two-form field strengths are coupled to S1-brane
and S0-brane solutions, and the corresponding charges and gauge
field configurations are discussed. In section 2.4 the R-symmetry
group of
 Lorentz transformations which leave the S-brane location fixed is discussed.
 This is argued to be broken,
possibly down to spatial rotations, at scales of order the
 brane thickness.  Section 3.1
embeds this discussion in string theory using the fact that the
worldvolume tachyon in an unstable D-brane (D-brane-anti-D-brane
pair) has a double well (Mexican hat) potential. In section 3.2
the discussion is generalized to S-branes within branes. Time
dependent D3-brane worldvolume configurations are described
(following \FarhiEH ) which have a nonzero value of $\int \tr
F\wedge F$ which is localized in time. These carry RR charge
corresponding to an SD(-1)-brane. Section 3.3 briefly mentions the
generalization to NS charges. In section 4 a description of these
S-branes as Dirichlet branes is considered. In section 4.1 the
boundary state for a D-brane with Dirichlet boundary conditions on
a timelike coordinate is constructed. Unlike the usual case,
the SD-brane boundary state has an on-shell component. This represents the
initial close strings which converge and create the SD-brane and the
outgoing closed strings which represent its decay products.
The corresponding classical
closed string field configuration  is
found to have singularities along the past and future light cones
of the brane. These arise (as for ordinary branes) because in the
leading order description the closed strings are squeezed into a
zero-width region along the brane. The RR charge associated to a
sphere surrounding the S-brane is computed from the boundary
state. In section 4.2 the problem of how the incoming closed
strings in the boundary state ``know'' to make an SD-brane is
discussed. We also present a puzzle concerning the brane
energy density, for which the soliton and D-brane pictures do not obviously
agree, and which indicates subtleties in the relationship between the
two. The worldvolume theory of the
SD-brane is also briefly discussed. Section 5 contains a
preliminary discussion of the classical (super)gravity solutions,
which might eventually provide holographic duals of the Euclidean
theories on the S-brane worldvolumes. As a warm up in section 5.1
we describe the S0-brane in D=4 Einstein-Maxwell gravity. In 5.2
we then find a solution corresponding to the SM5-brane in D=11.
These have singularities along the light cones. In section
6 we discuss the issue of the singularities and how they might
potentially be resolved. Finally section 7 concludes with a brief summary.

This work was in part inspired by the works of Hull \refs{\HullVG,\HullII}, who
proposed a so-called type II* string theory as the timelike T-dual
of type II string theory. He further argued that the II* theory
contains BPS, spacelike D-branes
holographically dual to de Sitter space.
The situation where all
coordinates obey Dirichlet boundary conditions (the SD(-1)-brane)
was discussed in the past
in order to introduce hard scattering behavior into string amplitudes
\refs{\GreenZV\GreenYT -\GreenIW} and define off-shell string amplitudes
\refs{\GreenWR,\CohenPV}.

Very recent interesting and possibly related
work on time-dependent string backgrounds includes
\refs{\vijay, \lms}.

\newsec{S-branes in D=4 Scalar Field Theory }
\subsec{An S2-brane Domain Wall}We start with a simple example of a
        scalar field with the
double well potential
\eqn\ffk{V(\phi)=(\phi^2-a^2)^2,}
with vacua at $\phi=\phi_\pm=\pm a$. We further weakly couple the scalar
field  to
some form of massless radiation. This theory has ordinary
2-brane domain wall solutions which interpolate between the two
vacua.

S2-brane solutions can also be constructed as follows.
Impose the initial data at $t=0$
\eqn\indt{\phi(\vec x,0)=0,~~~~~ \dot \phi(\vec x,0)=v,}
where $v$ is a small positive constant. This corresponds to
a scalar field about to fall off the local maximum at the origin.
As the system evolves forward in time, the scalar field will
dissipate energy into radiation, oscillating about and decaying towards
the $\phi_+$ vacuum.\foot{Strictly speaking in this case the full scalar plus
radiation
cannot fully settle down into its ground state because the S-brane fills space
and there is no asymptotic region  for the radiation to dissipate into.
We shall not dwell upon this feature because it
does not occur in the higher codimension branes discussed below which
are our real interest.}  On the other hand, since the initial conditions
are invariant under $\phi \to -\phi$ together with $t\to - t$,
evolving the system into the past leads to a similar configuration
asymptoting to the $\phi_-$ vacuum.

The full evolution of the system can be described as follows.
At $t=-\infty$, the scalar field is near $\phi=\phi_-$.
Finely tuned radiation conspires to excite the scalar field
and push it to the top of barrier. It then falls over the other side,
dissipating its energy into more radiation and decaying towards the $\phi_+$
vacuum. Because $\phi\to \pm a$ for $t \to \pm \infty$ (rather than
$x\to \pm \infty$), this is a
spacelike (rather than timelike) 2-brane.

One could also choose to specify initial data corresponding
to $\phi$ perched at its maximum along the null hypersurface
$t=x$ (together with vacuum initial data at $t+x \to -\infty,~~t>x$).
This leads to an N2-brane.

\subsec{An S1-brane Vortex} Now consider a complex scalar with the
Mexican hat potential
\eqn\dffk{V(\phi)=(\phi^*\phi-a^2)^2.}
This has the usual global cosmic string (timelike 1-brane) solutions.
S1-brane solutions arise from ${\it e.g.}$ the initial data
\eqn\tgh{\phi(x,y,z,0)=a\tanh z,~~~~~\dot \phi(x,y,z,0)=ive^{-z^2}.}
In the far future, after dissipating energy into radiation, the
scalar will settle down into a configuration with $|\phi|=a$,
but with $\phi=-a$ at $z=-\infty$ interpolating to $\phi=a$ at $z=\infty$
clockwise in the upper half plane. The symmetry of \tgh\ under
$t\to -t,~~\phi\to \phi^*$ then implies
that in the far past, the configuration originated in the vacuum with
$\phi=-a$ at $z=-\infty$ interpolating $\phi=a$ at $z=\infty$
counterclockwise in the lower  half plane.  The scalar field
makes one full circle about the origin as the asymptotic circle
in the $(z,t)$ plane is traversed.

This configuration can be described as finely tuned radiation which
comes in from $z=\pm \infty$, and then is fully absorbed by the scalar
field. This
excites the scalar field to the top of
the bump. It subsequently rolls down the other side of the
bump, emitting radiation as it goes.

This construction, as well as that of the previous subsection, have obvious
generalizations to both higher spacetime dimension and higher codimension.

\subsec{Charged S-branes}
We wish to consider the possibility of adding
adding axion charge to the 1-brane, associated to the
three-form field strength $H=dB$. This is accomplished via the coupling
\eqn\cplg{g \int B\wedge dj,}
where $g$ is a coupling constant and $j$ is the Klein-Gordon one form
\eqn\jdf{j={i \over 4 \pi a^2}(\phi^*\p_\mu \phi -\phi\p_\mu \phi^*)dx^\mu.}
For the timelike 1-brane, this gives an axion field strength
which falls of like $1 \over r$ away from the string.
The axion charge is
defined by the integral of $*H$ over a spatial contour encircling
the string.

The case of a spacelike 1-brane following from \tgh\
is less familiar.  On distances
large compared to the thickness of the brane, the axion field
has a source term given by
\eqn\fghm{d*H=g \delta^2(t^+,t^-) dt^- \wedge dt^+,}
where $t^\pm=t\pm z$. This equation is solved by
\eqn\foo{H=dx\wedge dy \wedge d\psi (t^+,t^-),}
where $\psi$ obeys the two-dimensional wave equation with
source
\eqn\ppf{2\p_+\p_-\psi = g\delta^2(t^+,t^-).}
In terms of advanced and retarded Green functions, a time symmetric
solution is
\eqn\psiso{\psi(t^+,t^-)={g\over 4 }\bigl(G_{\rm adv}(t^+,t^-;0,0)
+G_{\rm ret}(t^+,t^-;0,0)\bigr).}
Explicitly
\eqn\fhjp{\eqalign{ G_{\rm ret}(t^+,t^-;0,0)&=\Theta(t^+)\Theta(t^-),
\cr G_{\rm adv}(t^+,t^-;0,0)&=\Theta(-t^+)\Theta(-t^-), \cr
\psi(t^+,t^-)&={g\over 4}\Theta(t^+t^-),}}
$H$ has support on the light cone of
the S1-brane
\eqn\hsp{H={g\over 4}dx\wedge dy \wedge \bigl(dt^+\delta(t^+)\epsilon(t^-)+
dt^-\delta(t^-)\epsilon(t^+)\bigr) .}
It is easy to see that the integral of $*H$ around a contour encircling the
origin in the $(t^+,t^-)$ plane gives a nonzero charge.
Of course homogeneous solutions of \ppf\ (such as $G_{\rm adv} -G_{\rm ret}$)
may also be added to \fhjp.

This construction can also be generalized to other (co)dimensions.
For example, the S2-brane of section 2.1 can be coupled to a
4-form field strength $F=dC^{(3)}$ via the interaction
\eqn\imj{\int d\phi \wedge C^{(3)}.}
However the nature of the field strength around the brane depends
qualitatively on whether the codimension is even or odd. For example
consider an S0-brane in D=4 which has codimension three. This can
be coupled to
a Maxwell field, which obeys \eqn\ggj{dF=0,~~~d^\dagger F=dz\delta (t)\delta (x)\delta (y),}
corresponding to an S0-brane extended in the z-direction.
These equations have the solution
\eqn\slmn{F={\rm Re} \bigl( dz\wedge d{1 \over \sqrt
{t^2-x^2-y^2-i\epsilon}} \bigr)  .}
$F$ is real and supported inside the
past and future light cones of the brane. Note that this differs
from \hsp\ which is supported only on the light cone.  In general odd
(even) codimension solutions will be supported in (on) the light cone.

An interesting  physical realization of a charged S-brane can be
given as follows.\foot{We are grateful to N. Toumbas for this
example.} An electric current in a wire is purely spatial: there
is no $j_0$ component because the charge density vanishes. However
if the current persists indefinitely it is not localized in time
and so does not correspond to an S-brane. To get time localization
the current flow should stop and start. To arrange this,
consider initial data at
$t=0$ with no electromagnetic fields, and the electrons in the
wire all moving uniformly to the right. The electrons will
subsequently decelerate, and eventually come to rest,  due to
resistance in the wire and electromagnetic radiation produced by
their motion. Evolving backward in time from $t=0$, one arrives at
an initial state of incoming radiation with no current in the wire.
The full picture is
roughly that finely tuned incoming radiation momentarily excites a
current in the wire, which decays to outgoing radiation.
The charge measured by the integral of $*F$ over
the $S^2$ is given by the number of electrons which cross  any
three-dimensional ball whose boundary is the $S^2$.
 \subsec{R-symmetry}

In the last subsection, massless field solutions were constructed
on scales large compared to the S-brane thickness by solving a
wave equation with with delta function support on the brane. These
solutions are invariant under the transverse Lorentz
transformations which leave the location of the brane fixed. For an
Sp-brane in D spacetime dimensions,
this is an $SO(D-p-2,1)$ "R-symmetry".

However on shorter scales, the exact solutions do not have this
R-symmetry,\foot{Codimension one branes are an exception as the
symmetry is trivial.} as can be seen directly from the scalar
field initial data presented in section 2.2. Anything which is
localized to a nonzero finite width region in both space and time
can not have a Lorentzian symmetry (that would require
localization along the light cone), and there is no Lorentz
invariant regularization of the light cone singularity.\foot{We
are grateful to L. Motl for discussions on this point.} The best
exact symmetry one can hope for is $SO(D-p-2)$ spatial rotations.

It is conceivable that this might be describable as spontaneous
R-symmetry breaking in an S-brane field theory.

\newsec{S-branes in String Theory}

In this section we argue that spacelike p-branes appear in string theory.

\subsec{S-branes as Tachyon Solitons}
Let us start with an unstable D3-brane in IIA string theory.
This has a worldvolume tachyon $T$
with a potential roughly of the form \ffk.
A static unstable D3-brane corresponds to the tachyon perched
at the top of the potential. The theory at the bottom of the potential
is presumed to correspond to the closed string vacuum and have no
propagating open
string modes. The decay proceeds when the tachyon slips off
the top of the hill. The tachyon is coupled to closed string modes and as
it rolls off of the hill its energy is radiated off to infinity in outgoing
 closed string modes.

As argued by Sen \SenTT\SenMG, the stable D2-brane is a
tachyon kink solution
which interpolates between the two vacua ($T_\pm=\pm a$ in \ffk) on a
spacelike trajectory.
This has nonzero RR charge due to a threebrane worldvolume coupling \SenMG
\eqn\frt{\int dT \wedge C^{(3)}_{RR},}
as in \imj.
The S2-brane can then be constructed as a timelike rather than spacelike
kink. Impose initial data at $t=0$, exactly as in \indt\ with
$T =0$ and $\dot T>0$. The tachyon will roll off the hill towards
the $T_+$ vacuum. As it rolls the energy will be radiated off to infinity
in closed string modes and $T$ will approach $T_+$ in the infinite
future. Evolving to the past one has a time-reversed configuration in
which $T \to T_-$ in the infinite past. This results in an S2-brane
which carries RR charge due to the coupling \frt.

Sen also proposed an alternate description of the D2-brane as a
vortex in the D4-(anti-D4) system. This system contains a complex
tachyon with a Mexican hat potential. The RR field couples to the
vortex number via the coupling $\int C^{(3)}_{RR}\wedge dj$ where $j={\rm
Im}T^*dT$, the five dimensional analog of \cplg.  Hence the spacelike
vortex solutions of
section 2.2 become S2-branes carrying RR charge. A similar
construction on null initial surfaces gives N2-branes.

It is interesting to ask how ``thick'' the S-branes are in the time
direction. This corresponds to the decay time $t_{decay}$ of the
brane-antibrane pair
or the unstable brane. Naively one might guess that $t_{decay} \sim {1
\over g_s}$ in string units because the decay proceeds via
closed strings which couple with strength $g_s$.

We suspect that in fact this is not the case and that rather $t_{decay}
\sim {\cal O}(1)$. The reason is that the open string theory is
strongly coupled at the bottom of the potential. There are no perturbative
open string
oscillations whatever about the minimum - it is the closed string vacuum.
So it is hard to see how
$T$ can pass the minimum, let alone
undergo many oscillations before settling into vacuum. Rather we
suppose that $T$ reaches the minimum in a time of order the
string time, at which point all excitations can be described as some
kind of  closed string mode.

One might also imagine that the S-brane can be made arbitrarily
thick by making $v$ very small. However in this case quantum fluctuations
become important and push $T$ off of the top in
a time of order one in string units \ckl.

\subsec{S-branes within Branes as Evanescent Sphalerons}

Ordinary branes within branes can often be described in terms of smooth
gauge field configurations. For example a 0-brane in a pair of
4-branes can be described as an SU(2)
gauge instanton. There is a coupling $\int C^{(1)}_{RR}\wedge {\rm tr}F\wedge
F$, which implies
that the configuration carries the RR charge appropriate to a $D0$-brane.

We wish to argue that S-branes within branes can have a similar
description.
While the relevant solutions are more complicated than the tachyon
kinks of the previous subsection, the discussion is not plagued by the
uncertainties in the tachyon effective Lagrangian.
In particular we consider the S(-1)-brane as an event in the  D3-brane.
What we need is a Lorentzian configuration with a nonzero integral of
${\rm tr}F\wedge F$ over some localized spacetime region. We can't take
$F=*F$, since that has no real solutions in Minkowski space.
Let us take the D3-branes to be slightly separated so that the gauge
symmetry is spontaneously broken. Then it is known that there are
noncontractible
loops in the
three-dimensional configuration space which begin and end at the vacuum \taub.
However the initial and final vacua have different winding numbers and
correspondingly the sequence of configurations has a nonzero value of
$\int {\rm tr}F\wedge F$. This path passes through a saddle point
where the energy takes it maximum value. This saddle point is a static but
unstable solution of the equations of motion known as a sphaleron
\MantonND\KlinkhamerDI.

Now let us specify initial data at $t=0$ for a Lorentzian solution
corresponding to
sphaleron with a small initial velocity pushing it toward the higher
winding vacuum. As time evolves to the future, the sphaleron slides
down the hill, energy
is radiated off the brane into closed string modes, and the gauge
field settles into
its vacuum state. Going backwards in time, the system approaches the
lower-winding vacuum. Altogether the spacetime configuration traverses
the noncontractible loop in the configuration space and has integral
value of ${\rm tr}F\wedge F$.\foot{A
cogent discussion of this type of configuration is given in \FarhiEH.
Our system is however a bit simpler in that the leakage of energy off of the
brane allows the gauge field to fully approach the vacuum asymptotically.}
The worldvolume coupling  $\int C^{(0)}_{RR}{\rm tr}F\wedge F$ then
implies that this Lorentzian configuration carries the same RR
charge as a Euclidean D-instanton.

When the branes are coincident the gauge symmetry is not broken and such
configurations still exist. However it appears that their energy can
be made arbitrarily small by
spreading them over a large region of space.

We wish to emphasize that these configurations are in no sense the analytic
continuation of the usual SU(2) instanton solutions. These latter
have complex values of the fields and are not admissible
as real solutions in Minkowski space.

The $g_s \to 0$ limit of these configurations is rather
different than that of the tachyon kink S-branes. At $g_s = 0$
the Yang-Mills-Higgs theory on the D-brane completely decouples from
the closed string sector. One then has scattering configurations
with nonzero ${\rm tr}F\wedge F$ of the
type discussed \FarhiEH\ which will source the RR charge.
The question of exactly where the RR charge is localized
is then rather subtle although likely largely answered in the work of
\FarhiEH.  We will not try to address it herein.

\subsec{NS S-branes}

Duality relates all types of branes and hence suggests that branes
carrying NS charge can
also be spacelike. In fact the type I /heterotic 5-brane can be
described as an instanton in $SO(32)$ or
$E8$. As such its spacelike configuration might be described by the
evanescent sphalerons of the preceding subsection. However here,
because there is no extra region for the energy to leak off into,  we
encounter the subtleties described in \FarhiEH.

 \newsec{Spacelike D-branes}
  In the previous section time-dependent solutions of string theory
corresponding to spacelike branes were described. They are
spacelike counterparts of Sen's solitonic description of the usual
D-branes  and in particular carry the same RR charges. It is
natural to suppose that in string perturbation theory they can be
described by an SD-brane - $i.e.$ a
conformal field theory in which the timelike
worldsheet coordinate $X^0$, as well as $8-p$ spacelike coordinates,
obey Dirichlet boundary conditions. In the null case one takes
$X^+$ to obey Dirichlet boundary conditions, but we defer this
case to later consideration.\foot{Interestingly, ND-branes can
have on-shell propagating open strings on their  worldvolumes.}
\subsec{The Boundary State}
The SD-brane acts as a source for massless and massive closed string fields.
As is the case for timelike D-branes, these fields can be succinctly described
at leading order in $g_s$ with the use of a boundary state. The
spacelike case
will turn out to differ only in a few signs from the timelike case.
We refer the reader
to \DiVecchiaPR, whose conventions we follow, for details of the latter.
The bosonic part of a boundary state which imposes the boundary
conditions of a spacelike $p$ brane is given by imposing Dirichlet boundary
condition $X^a=y^a$
on the first $9-p$ coordinates $X^a, a=0,1,\cdots, 8-p$
and Neumann boundary conditions on
the last $p+1$ coordinates $X^i,i=9-p,\cdots,9$.
\eqn\bounst{\eqalign{(\alpha_{n}^\mu +O^\mu_{~~\nu} \bar \alpha_{-n}^\nu)\mid
B,\eta\rangle &=0, \quad \quad \mu={0,1,\cdots, 9}\cr (\psi_r^\mu - i
\eta O^\mu_{~~\nu}  \bar{\psi}^\nu_{-r})\mid B ,\eta\rangle&=0, \quad \quad
\mu={0,1,\cdots, 9}\cr
 (q^a - y^a) \mid B,\eta\rangle &=0, \quad \quad a=
0,1,\cdots,8-p \cr \quad \quad p^i \mid B,\eta \rangle&=0, \quad
\quad  i=9-p,\cdots,9}} where
$O^{\mu\nu}=diag(-\eta^{ab},\delta^{ij})$ and $\eta =\pm 1$. The
solution of these equations is given by
\eqn\zmpart{\mid B,\eta
\rangle=N_p \exp\big( - \sum_n {1\over n} O_{\mu\nu}
\alpha_{-n}^\mu \bar \alpha_{-n}^\nu+ i\eta \sum_{r} O_{\mu\nu}\psi_{-r}^\nu
\bar \psi_{-r}^\nu\big) \int d^{p+1} k e^{i k_a y^a}   \mid 0,\eta
\rangle, }
Here we did not display the ghost part of the boundary state which is the
same as for the standard timelike branes, and both the ground state and
  oscillator moding differ between
R-R and NS-NS sectors.
The zero mode part in the NS-NS and R-R
sector are given by
\eqn\zmod{\eqalign{\mid 0, \eta \rangle_{NS}&=-
i \mid k^a, k^i=0\rangle,\cr \quad \mid 0, \eta \rangle_{R}
&=\Big(C\Gamma^{9-p\cdots 9} {1+i \eta \Gamma^{11}\over 1+i
\eta}\Big)_{\alpha\beta}\mid \alpha \otimes \bar \beta, k^a,
k^i=0\rangle.} }
We assume that the normalization $N_p$ is the same as in the
timelike case, namely
\eqn\np{N_p=2^{2-p}\pi^{{7 \over 2}-p}{\alpha'}^{3-p \over 2}.}
The D-brane boundary states are then given by the
GSO projection invariant combinations $\mid
B,\eta=+1\rangle_{NS}-\mid B,\eta=-1\rangle_{NS}$ and $ \mid
B,\eta=+1\rangle_{R}+\mid B,\eta=-1\rangle_{R}$ for the NS-NS and
R-R sector respectively.

A state in the first quantized Hilbert space of a string can
be interpreted as a classical
string field configuration describing massive and
massless closed string fields in spacetime. A solution $|\Phi>$ of the
linearized equation is given in terms of the boundary state
via the relation\GreenUM\DiVecchiaPR
\eqn\dffc{|\Phi>={1 \over L_0+ \bar L_0-a} |B>,}
where $a=1,0$ in the NS-NS and R-R sector respectively.
\dffc\ is ambiguous because the states can be on shell and the
denominator vanish. This is the expected ambiguity corresponding to the
freedom to add homogeneous solutions of the equations of motion.
For the closed string fields
the overlap of the boundary state \zmpart\ can be related to the
corresponding supergravity solution. For example the NS-NS dilaton, in the
normalization of \DiVecchiaPR\ obeys
\eqn\subo{\partial_a\partial^a  \phi(x) = \langle \phi(x) \mid B (y)
\rangle ={3-p \over \sqrt{2} }N_p \delta^{p+1}
(x-y).}
The (p+1)-form R-R potential  is of special interest. In the gauge
$d*C_{p+1}=0$ it obeys
\eqn\malla{ \partial_a \partial^ a C_{9-p \cdots 9}(x)
= 4N_p \delta^{p+1}(x-y).}
Hence for the spacelike branes the small fluctuations around flat
space will be solutions of the wave equation with a delta function
source in the directions transverse to the worldvolume.  The equations
\subo\ and \malla\ have the same numerical coefficients as
their timelike D-brane counterparts.  For a
time symmetric solution we use the advanced plus retarded
propagator, as in section 2.3.
This will be supported within the past and future
light cones of the brane. We note that the fields diverge not just
at a point in the transverse space (the location of the brane) as
in the case of timelike branes but instead along the lightcone in
transverse space. This divergence will be further discussed in sections
4.2 and 6.

The charge of the brane can be defined by an integral
of the  RR field strength or its dual $*F_{p+2}=dC_{p+1}$ over an 8-p sphere
surrounding the brane. This 8-p sphere has both timelike and spacelike
components. Evaluating this charge  with  Gauss's law and
using
\malla, we find
\eqn\dfll{\int_{S^{8-p}}*F_{p+2}=4N_p     .}
This is numerically
the same result obtained for the charges of ordinary branes defined by
integrals over spatial spheres.
It also agrees with the charges found in the solitonic description of the
branes. This result is the main reason for
expecting the solitonic and Dirichlet
S-branes represent the same basic
object.
\subsec{Discussion}
Schematically, the linearized closed string field configuration
is of the form
\eqn\drt{\Phi \sim G_{\rm adv}+G_{\rm ret},}
and has a delta function source on the S-brane worldvolume.
Consider the alternate configuration
\eqn\drt{\tilde \Phi \sim G_{\rm adv}-G_{\rm ret}.}
This obeys the linearized field equations $without$ a delta
function source anywhere. Furthermore $\tilde \Phi$ agrees $exactly$
with $\Phi$ in the past of the brane.

This raises an important question: how do the incoming closed strings know
whether or not to make an S-brane? A possible answer to this question can
be found in the tachyon soliton description of the S-brane.
This crucially involved the open string tachyon field. However
this field cannot be seen in any order in perturbation theory about the
closed string vacuum which sits at the minimum of the tachyon potential.
The open strings are hidden
non-perturbative degrees of freedom in the closed string vacuum.
Therefore we would not have expected to see them in a perturbative
description of the incoming state as an excitation
of the closed string vacuum. In order to distinguish those
incoming state which do and do not form S-branes, we need a much finer
description than that supplied by the boundary state.

A second issue concerns the R-symmetry . The boundary state is
invariant under the $SO(8-p,1)$ Lorentz transformations
transverse to the branes. As discussed in section 2.4, this cannot
be an exact symmetry on scales of order the brane thickness, which
we expect to be $\sqrt{\alpha '}$ times a power of the string
coupling. Hence higher corrections should not preserve this
symmetry.

This observation is related to brane energy.\foot{We are grateful to
G. Horowitz and E. Verlinde for discussions on this topic.}
The stress energy density of the
brane itself has nonzero components \eqn\stre{T_{ij} \sim {1 \over
g_s}\delta_{ij}\prod_{a=0}^p\delta(x^a-q^a).} This form is largely
required by the R-symmetry of the leading order description.  It
is easy to check that it is conserved. \stre\ gives a positive
energy density in any Lorentz frame but does not obey the dominant
energy condition. \stre\ seems to be in contradiction with the
tachyon picture, which suggests a nonzero $T_{00}$ of order ${1
\over g_s}$. However there is an additional contribution to the
energy from the closed strings in the boundary state. This gives
an energy density which diverges on the light cone. The origin of
this divergence is that the boundary state tries to confine
strings to a point. A similar divergence in the Coulomb-like
energy occurs for ordinary timelike D-branes, and is resolved by
higher corrections.  In principle the discrepancy in $T_{00}$ in
the tachyon soliton and the D-brane description of S-branes could
also be resolved by higher corrections which impart a finite width
to the brane. Hence the resolution of this singularity should be
seen at the same time as the breaking of the R-symmetry. We have
not understood how this would work in detail.

We expect that the worldvolume theory will be a kind of
Euclideanization of the usual D-brane theory, including a minus
sign in the kinetic term for the field corresponding to
fluctuations in the $X^0$ directions. The $R$-symmetry (which
should be a symmetry of the action) will be the group of
transverse rotations which is $SO(9-p-1,1)$ rather than
$SO(10-p-1)$.

\newsec{Spacetime Solutions}

In this section we discuss several solutions of the supergravity
equations corresponding to S-branes with odd codimension. In
principle these might provide holographic duals of the SD-brane
field theories.  The basic idea pursued here is simply to consider
a general ansatz with $ISO(p+1)\times SO(D-p-2,1)$ symmetry.   Of
course, to to account for spontaneously broken R-symmetry, a less
restrictive ansatze is appropriate. (AdS/CFT examples with
spontaneously broken R-symmetry are described in \manu.) We hope
to give a more complete discussion of the solutions and their
properties in future work.

As there is no supersymmetry the solutions are considerably more
complex than their timelike partners.\foot{We note that naive
analytic continuation of the familiar timelike p-brane solutions
leads to a complex field strength.} The even codimension case is
qualitatively different because at linear order the massless
fields are supported on (rather than in) the S-brane light cone.
\subsec{An S0-brane in D=4} In this subsection we consider a
simple solution of D=4 Einstein-Maxwell gravity corresponding to a
charged S0-brane. This could be relevant to a string
compactification with suppressed scalar moduli. The
Einstein-Maxwell action is \eqn\eact{\int d^4x\sqrt{-g}
\bigl(R-F^2\bigr).} In the absence of gravity, the solution
corresponding to a charged S0-brane was given in \slmn, and has an
$SO(2,1)\times R$ symmetry. It is natural therefore, when
including the gravitational back reaction, to impose this
symmetry. We accordingly consider the ansatz
\eqn\ansm{ds^2=-{d\t^2 \over \l^2}+\l^2dz^2+R^2dH_2^2,}
\eqn\ansf{F=Q\epsilon_2,} where $R$ and $\l$ are functions of $\t$
only and $dH_2^2$ ($\epsilon_2$)  is the line (volume) element on
the unit H$_2$ with curvature \eqn\crv{R_{ab}=-g_{ab},} for
$a,b=0,1$. It is easy to check that \ansf\ obeys the Maxwell
equations $dF=d*F=0$ as well as \eqn\frg{F^2={2 Q^2 \over R^4}.}
It remains to determine $R$ and $\l$ from the Einstein equations.
One finds \eqn\fdy{\eqalign{R_{ab}&={g_{ab}\over
R^2}\bigl((RR'\l^2)'-1\bigr)\cr &=\half g_{ab}F^2,\cr
R_{zz}&={g_{zz}\over R^2}(R^2\l \l')'\cr &=-\half g_{zz}F^2,\cr
R_{\tau\tau}&={g_{\tau\tau}\over R^2} \bigl( R^2 ( \l \l')' + 2 R
\l ( \l R')'\bigr) \cr &=-\half g_{\tau\tau }F^2, }} where $'$
denotes differentiation with respect to $\t$. Equating the two
expressions for $F^2$ gives \eqn\iof{(R^2\l^2)''=2.} Integrating
twice gives \eqn\faz{R^2={\t^2-\t_0^2 \over \l^2}.} $\t_0^2$ is a
(possibly negative) integration constant and we have suppressed
the second integration constant corresponding to shifts of $\t$.
Substituting \faz\ and \frg\ into the second equation of \fdy\
yields \eqn\kkl{\bigl((\t^2-\t_0^2)(\ln \l^2)'\bigr)'=-{2 Q^2\l^2
\over \t^2-\t_0^2}.} Defining the new variable $t$ by
\eqn\xt{\t=-\t_0\coth {\t_0 t},} so that
\eqn\hji{(\t^2-\t_0^2)\p_\t= \p_t.} Then \kkl\ becomes simply
\eqn\jij{\p_t^2\ln \l^2=-2 Q^2\l^2.} The solution is
\eqn\sxz{\l^2={ \alpha^2 \over Q^2} {\rm sech}^2 \alpha t,} where
$\alpha $ is an integration constant. The third equation of \fdy\
then requires $ \alpha= \tau_0$, and hence \eqn\sxxz{\l^2={
\tau_0^2 \over Q^2} {\tau^2-\tau_0^2 \over
\tau^2},~~~~~~R^2={Q^2\tau^2 \over \tau_0^2}     }
The line element
is
\eqn\asx{ds^2=- {Q^2 \over  \tau_0^2 } {\tau^2\over
\tau^2-\tau_0^2}      d\tau^2+  { \tau_0^2 \over Q^2}
{\tau^2-\tau_0^2 \over
\tau^2}    dz^2
+ {Q^2\tau^2 \over \tau_0^2}   dH_2^2.}
The asymptotic region with $R\to \infty$ is at $\tau\to\pm\infty$ and
the metric behaves as
\eqn\pgf{ ds^2\to-{Q^2 \over \tau_0^2}d\tau^2+{\tau_0^2 \over Q^2}dz^2
+{Q^2 \tau^2
\over \tau_0^2}dH_2^2,}
and is asymptotically locally flat. Near the
brane at $\tau\to \tau_0 $ one has, defining  $t^2= \tau^2- \tau_0^2$
\eqn\fyyu{ds^2\to-{ Q^2 \over \t_0^2}dt^2
+{\t_0 t^2 \over Q^2} dz^2
+{2Q^2}dH_2^2.}

\subsec{An S5-brane in D=11}
The bosonic
action of eleven dimensional supergravity is given by (dropping the WZ
term which will not be important in the following).
\eqn\acaa{S=\int d^{11}x \sqrt{-g}\big( R-{1\over 2\cdot 4 !}F^2\big)}
The equation of motion is given by
\eqn\eqom{R_{\mu\nu}= {1\over 2 \cdot 3!} F_{\mu\lambda\rho \sigma}
F_{\nu}^{~~\lambda\rho \sigma}- {1\over 6 \cdot 4!}
  g_{\mu\nu} F^2.}
An ansatz \foot{Similar solutions were obtained in
\LuJK\LuER\LukasEE\LukasIQ\Quevedo.} which simplifies the equations of
motion is \eqn\metans{ds^2 = -e^{-4f + 8g}dt^2 + e^{2f} dx_6^2+
e^{2 g-4f}dH_4^2, } where $dH_4$ is the metric on a hyperbolic
space with  negative curvature, and \eqn\fsan{* F = h(t) dt\wedge
dx_1\wedge\cdots \wedge dx_6 .} Solving $d F=0$ leads to
\eqn\hsol{h(t)= e^{12 f} q .} With this ansatz \eqom\ becomes
\eqn\eqomb{\eqalign{f''+{q^2\over 6} e^{12 f}&=0,\cr g''- 3 e^{6g}
&=0,\cr -18 f'f'+ 12 g'g' + 2 f'' - 4 g'' - { q^2\over 6} e^{12
f}&=0. }} It is easy to see that the system of equations \eqomb\
is
 equivalent to the following two first
oder equations.
\eqn\sola{g'g'- e^{6 g}=c, \quad \quad f'f'+{q^2 \over 36} e^{12 f}
={2 c\over 3},}
where $c$ is a positive integration constant.
The solutions of \sola\ are easily obtained by integration
\eqn\solb{\eqalign{f(t)&= {1\over 12} \ln\left( { c \over \cosh^2 \big(2
\sqrt{ 6c}(t-t_0)\big)}\right)-{1\over 12}\ln ( q^2/24), \cr
g(t)&= {1\over 6} \ln \left( {c \over \sinh^2\big( 3
\sqrt{c}(t-t_1)\big)}\right).}}
The integration constant $c$ can be eliminated by rescaling $t\to
t/\sqrt{c}$ and $ x_i  \to x_i/ c^{1\over 12}$. Furthermore we are free to set
$t_1=0$ by a shift. This leaves one constant $t_0$. The metric takes the form
\eqn\metc{\eqalign{ds^2&= -\left( {q^2 \over  24} \right)^{{1\over 3}}{
\big( \cosh
\sqrt{24}(t-t_0)\big)^{2\over 3}\over \big(\sinh 3 t\big)^{8\over 3} }
\left(-dt^2+(\sinh 3 t)^2  dH_4^2 \right)
\cr &\;\;\;+\left( {q^2 \over  24} \right)^{-{1\over 6}} {1\over \big( \cosh
\sqrt{24}(t-t_0)\big)^{1\over 3}} dx_6^2. }}
The asymptotic region is at $t\to 0$ where the radius of the $H_4$
diverges.
Defining $u=t^{-1/3}$, near $t=0,~u=\infty$ the metric becomes
\eqn\limeta{\eqalign{ ds^2_{t\to0}&\sim - \left( {q^2 \over  216}
\right)^{{1\over 3}}{(\cosh \sqrt{24} t_0)}\left(-du^2+u^2dH_4^2\right)
+ \left( {q^2 \over  24}
\right)^{-{1\over 6 }}{1\over (\cosh \sqrt{24} t_0)^{1\over 3}}
dx_6^2\cr
&\;\; ,  }}
which is locally flat space in Rindler coordinates.
The Ricci scalar behaves as
\eqn\rics{R= 8\left({3\over q^2}\right)^{1\over 3} \left( \sinh( 3
t)\over \cosh(2 \sqrt{6 }(t-t_0)) \right)^{8\over 3} .}
The large $t$, near-brane behavior is given by
\eqn\lilt{\eqalign{ds^2 \sim& -  \left( {q^2  \over  24}\right)^{{1\over
3}}\exp\Big( (-8+4\sqrt{2\over 3})t -4\sqrt{2\over 3} t_0\Big)dt^2+\left(
{q^2  \over 24}\right)^{-{1\over 6}}
\exp\Big( - 2\sqrt{2  \over 3} (t-t_0)\Big)dx_6^2\cr
&+\left( {q^2 \over   24}\right)^{{1\over
3}} \exp\Big( ( 4\sqrt{2\over 3}-2) t -  4\sqrt{2\over 3}  t_0\Big) dH_4^2
+{\cal O}(1).}}
Even though the Ricci scalar tends to zero in this region the geometry is
singular because for example the coefficient of $dx_6^2$ vanishes.

\newsec{Singularities}

Singularities have been encountered in several related places
in our discussion. In constructing the profile of the linearized
massless fields
sourced by an S-brane, one finds singularities along the
light cone of the brane. This divides the spacetime into three regions:
the interior of the future brane lightcone,  the interior of
the past brane lightcone, and the spacelike separated region.
Related singularities arose in construction of the full supergravity
solutions in section 5.

We envision four possible fates of these singularities in the full theory:

\noindent {\bf (i)}  The singularity is smoothed out by stringy effects
which are nonperturbative in $\alpha'$. This is the case for the
Coulomb-like singularities of ordinary
D-branes. It is suggested by the
tachyon soliton picture, in which the source for the various
closed string fields are spread out over a region of size of order
$\sqrt\alpha'$. As discussed in 2.4 and 4.2 this resolution should be
accompanied by breaking of the Lorentzian R-symmetry.

\noindent {\bf (ii)}  The singularity is smoothed out by stringy
effects which are nonperturbative in $g_s$.

\noindent{\bf (iii)} The maximal extension of the
future and past lightcone regions are
geodesically complete but do not include one another or the
spacelike region. Something like this occurs for example
for the ordinary D3 solutions. It also occurs for
Hull's  SD3-solutions of the II* theory, for which the near
brane region is dS$_5$.

\noindent {\bf (iv)} The singularities are intrinsically ``bad''
and the S-brane configurations herein should not be considered as
fully consistent solutions of string theory.

\newsec{Summary}
   We have shown that ordinary field theories contain time-dependent
topological solutions corresponding to S-branes, or branes with only
spatial dimensions. S-branes can carry charges measured by
flux integrals over spheres with both space and time dimensions.
It follows from
Sen's description of the tachyon action of unstable D-branes that
such objects exist in string theory. Furthermore they carry
the charges expected of an S-Dirichlet brane, which has a Dirichlet
boundary condition in a timelike direction.  Light cone singularities
are encountered in both the perturbative string and supergravity
descriptions of S-branes, which are analogs of the short-distance singularities
of ordinary D-branes. The resolutions of these singularities
must be understood before the proper role of S-branes in string theory can
be determined.

{\bf Acknowledgments.} It is a pleasure to thank G. Horowitz,
S. Minwalla, L. Motl, J. Polchinski, N. Seiberg, S. Shenker, N. Toumbas,
 C. Vafa and E. Verlinde
for useful discussions. This work was supported in part by DOE
grant DE-FG02-91ER40655.

\listrefs
\end